\documentclass{aa}  
\usepackage{graphicx}
\def\la{\raise.5ex\hbox{$<$}\kern-.8em\lower 1mm\hbox{$\sim$}}
\def\ma{\raise.5ex\hbox{$>$}\kern-.8em\lower 1mm\hbox{$\sim$}}

\def\kms{$\rm km\, s^{-1}$}
\def\cm3{$\rm cm^{-3}$}
\def\Ts{$\rm T_{*}$}
\def\Vs{$\rm V_{s}$}
\def\n0{$\rm n_{0}$}
\def\B0{$\rm B_{0}$}

\def\erg{$\rm erg\, cm^{-2}\, s^{-1}$}
\def\mum{$\mu$m}
\def\agr{a$_{gr}$}

\def\Hb{H$\beta$}
\usepackage{txfonts}
\begin{document}
   \title{The symbiotic star H1-36}

   \subtitle{A composite model of line and continuum spectra from radio to ultraviolet}

   \author{R. Angeloni
          \inst{1,2},
          M. Contini\inst{2,1},
	  S. Ciroi\inst{1}
          \and
          P. Rafanelli\inst{1}
          }

   \offprints{R. Angeloni}

   \institute{Dipartimento di Astronomia, Universit\`a di Padova, Vicolo
dell'Osservatorio 2, I-35122 Padova, Italy\\
              \email{rodolfo.angeloni@unipd.it, stefano.ciroi@unipd.it, piero.rafanelli@unipd.it}
         \and
             School of Physics and Astronomy, Raymond and Beverly Sackler Faculty of Exact Sciences, \\Tel Aviv University, Tel Aviv
69978, Israel\\
             \email{contini@post.tau.ac.il}}

   \date{Received - ; accepted -}

% \abstract{}{}{}{}{} 
% 5 {} token are mandatory
 
  \abstract
  % context heading (optional)
   %{}
   {In this paper we analyse the spectra of D-type SS H1-36 within a colliding-wind scenario.}
  % aims heading (mandatory)
   {We aim to analyse the properties of this object taking into account the observational data along the whole electromagnetic spectrum, in order to derive a self-consistent picture able to interpret the nature of the system as a whole.}
  % methods heading (mandatory)
   {After constraining the relative physical conditions by modelling more than 40 emission lines from radio to UV, we are able to explain the continuum spectral energy distribution by taking into account all the emitting contributions arising from both the stars, the dust shells and the gaseous nebul\ae{}.}
  % results heading (mandatory)
   {A comprehensive model of the radio spectra allows to reproduce the different slopes of the radio profile and the turnover frequency, as well as the different size of the observed shocked envelope at different frequencies in the light of the different contributions from the expanding and reverse nebul\ae{}. The IR continuum unveils the presence of two dust shells with different radii and temperatures, which might be a distinctive feature of D-type symbiotic systems as a class of objects. The broad profiles of IR lines direct us to investigate whether an X-ray jet may be present.}
  % conclusions heading (optional), leave it empty if necessary 
   {This insight leads us to indicate H1-36 as a promising X-ray target and to encourage observations and studies which consistently take into account the complex nature of symbiotic stars throughout the whole electromagnetic spectrum.}

   \authorrunning{Angeloni et al.}
   \titlerunning{The symbiotic star H1-36}
   \maketitle
%
%________________________________________________________________
\section{Introduction}
Symbiotic systems (SS) are currently understood as interacting binaries composed of
a compact star, generally but not necessarily a white dwarf (WD) which is the source of the ionizing radiation; a cool giant star, which is at the origin of dust formation and ejection episodes; and different emitting gas and dust nebul\ae{}. The large variety of the properties of these systems often makes their classification as a single group difficult; however, these differences are probably more related to distinct physical conditions like masses, orbital periods and temperature of the hot source, than different kind of phenomena. 

Because of the associated circumstellar matter and energetic activity, many SS emit detectable radiation across nearly the entire electromagnetic spectrum. On the basis of near-infrared (NIR) colours, SS were classified in S and D types (Webster \& Allen 1975) according to whether the cool star (S-type) or dust (D-type) dominates the 1-4 \mum \, spectral range. As with regards to the continuum radio emission, it has been modelled as thermal bremsstrahlung radiation arising from (a) the photoionized component of the cool star wind (Seaquist et al. 1984, hereafter STB - Taylor \& Seaquist 1984), (b) a wind associated with a slow nova-like eruption from the hot companion, (c) the interaction region involving both winds in mutual collisions (Watson et al. 2000, Kenny \& Taylor 2005, Bisikalo et al. 2006). Whatever the case, it is widely recognised that the radio spectra provide unique informations about the mass-loss phenomena at the basis of symbiotic activity. Eventually, there is growing evidence that SS can be the site for jet structures and synchrotron emission, confirming the primary role of shocks in the nowadays widely accepted interpretation of SS as colliding-wind binary systems.\\

H1-36 was among the most interesting objects in the extensive Purton's radio survey of emission-line stars (1982). The particular attention that was drawn to it since the Haro's discovery (1952) was due to a curious set of misunderstood and genuine insights: catalogued as planetary nebula despite its extremely high-excitation emission line spectrum and its imposing infrared excess, H1-36 was even thought to be, now we know erroneously, the optical counterpart of the \textit{Uhuru} hard X-ray source 3U 1746-37 (Giacconi et al. 1974). Finally, Allen permanently classified H1-36 as a D-type SS by including it in his catalogue (Allen 1981) and dedicating to it a paper (Allen 1983) that so far still represents the most complete observational work on such object.

One of the most important results emerging from that study is that the Mira star experiences an extraordinary reddening ($A_v \sim 20$ mag.), especially when compared to the one towards the emission-line regions (only 2.2 mag.): this allowed Allen to point out that the cool component is heavily embedded in its own circumstellar dust shell, illuminated from the outside by a $T_* \sim$ 150.000 K star. The derived cool star and dust temperatures would then be, respectively, $T_M \sim$ 2500 K and $T_d \sim$ 700-800 K.\\
Another interesting feature of H1-36 lies in the radio range: as a matter of fact, it is one of a small number of radio sources whose spectra are flat at high frequencies but turn towards a spectral index near +1 at lower frequencies (Purton 1977). Several papers hence tried to explain this behaviour in terms of free-free emission from the ionized cool star wind, unfortunately without taking into account the possibility of any outflow from the compact star (for a further description of this models, see Sect. 4.2.1).\\
The following papers, which presented both imaging (e.g. Bhatt \& Sagar 1991; Corradi et al. 1999, who resolved the complex nebula at radio and optical wavelengths) and new spectra (e.g. Costa \& de Freitas Pacheco 1994; Pereira 1995; Pereira et al. 1998), have not substantially been modifying the original scenario drawn by Allen's analysis, whose physical parameters have still to be considered as the most reliable ones available in literature. Interestingly, H1-36 is the only SS known to support \textit{OH-} (Ivison et al. 1995), as well as \textit{SiO-} (Allen 1989) and $H_2O$- masers (Ivison et al. 1998).\\

In this paper we aim to model H1-36 (Table \ref{tab:ID}) by combining literature observations from radio to UV collected over a period of about 30 years (Sect. 2). Within a colliding-wind scenario and by means of the SUMA code (Sect. 3), we start modelling more than 40 emission lines from IR to UV in order to constrain the physical conditions across the whole system (shocked nebul\ae{}, dust shells, hot and Mira stars). In the light of the derived parameters, we are then able to fit the composite continuum SED in a self-consistent way. The results of this cross-checked method are extensively presented in Sect. 4. Concluding remarks appear in Sect. 5.
\section{Observational details}
\subsection{Radio-mm observations}
In Table \ref{tab:radioref} we present the data we collected from literature in order to ensure as much a complete spectral coverage in the radio-mm range as possible.\\
The most important references about the H1-36 radio spectrum remain Purton's papers (Purton 1977, 1982), in which observations from about 2 GHz to 90 GHz obtained with the CSIRO 64-m telescope at Parkes, Australia, the NRAO 11-m telescope at Kitt Peak, USA, and the NRAO three-element interferometer at Green Bank, USA, are presented. In the next years several works, focused mainly on the mass loss properties of the cool stellar component (e.g. Jones 1985) and on the presumed correlations among different spectral bands (Seaquist et al. 1993), performed observations at intermediate frequencies. They allowed us to extract a well defined radio profile and eventually to confirm that the object has not been significantly variable in the last 30 years. \\ 
This ensures that collecting data coming from a wide temporal range is, at least for this SS, a reliable approach. Summarising, we are able to present a radio-mm profile composed by 16 points from 843MHz to 230GHz (Fig. \ref{fig:var}, bottom panel).
\begin{table*}[!ht]
\centering \caption{The main parameters of H1-36 D-type symbiotic star.\label{tab:ID}}
\begin{tabular}{ccc}\\ 
\hline  \hline
- & - & Ref.\\
\hline
$\alpha$ (J2000)& 17 49 48.1 &Belczy{\'n}ski et al. (2000) \\
$\delta$ (J2000) & -37 01 27.9 &Belczy{\'n}ski et al. (2000)\\
WD temperature & 1.5 10$^5$ K & Allen (1983)\\
Cool star spectrum & Mira - M4-M5& Allen 83 - Medina-Tanco \& Steiner (1995)\\
Mira temperature & 2500 K & Allen (1983)\\
Mira pulsation period & 450-500 d & Whitelock (1987)\\
Dust temperature & 700 K & Allen (1983)\\
Distance & 4.5 kpc & Allen (1983)\\
Binary separation & 3 10$^{16}$ cm & Allen (1983)\\
Orbital period & $\geq$10 D$^{3/2}$ yr& Allen (1983)\\
Other names & PK 353-04 1=Hen2-289= & Belczy{\'n}ski et al. (2000)\\
 & =Haro 2-36=IRAS 17463-3700 & \\
\hline
\end{tabular}
\end{table*}
\subsection{Infrared data}
\begin{figure}[!hb]
\begin{center}
\includegraphics[width=0.5\textwidth]{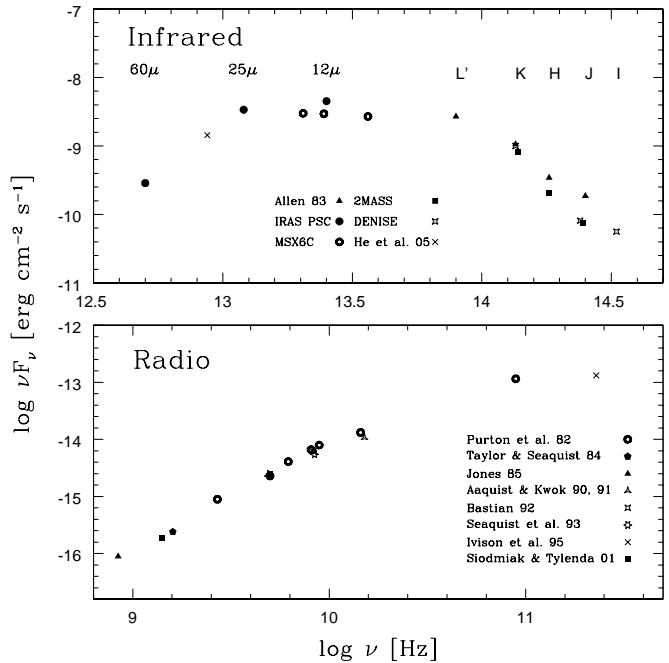}
\caption{Observational data. Top panel: IR spectral range - for sake of clarity we don't show the ISO spectrum. Bottom panel: radio spectral range.\label{fig:var}}
\end{center}
\end{figure}

D-type SS are known to be variable systems in the IR range, where we observe the contributions of different emitting components, i.e. cool star and dust shells. Quite surprisingly, as in the radio case, the H1-36 IR spectra do not show dramatic variations in the data flux over the years we are going to discuss (Table \ref{tab:irref}, Fig. \ref{fig:var}, top panel): indeed, IRAS points (1982) agree well with ISO-SWS spectrum (taken in 1996) and with the 34.6 \mum\ measurement presented in He et al. (2005); also the accordance with the MSX6C photometric points (Egan et al. 2003) is reasonably good. At NIR wavelengths the Allen's NIR bands are slightly shifted with respect to the DENISE and 2MASS bands, especially the H and J bands which feel the increasing contribution of the cool star. This is confidently due to the Mira variations, whose period is about 450-500 days (Whitelock 87).
\subsection{Optical spectra and UV spectrophotometry}
Also in this case the analysed data are a collection of different observing sessions from 1975 to 1979. These are presented in Allen (1983), where the reader is sent for further technical details. Moreover, in that paper it is explicitly stated that the optical data "are combined in the belief that variability of the lines is slight". 
This has been even more reconfirmed by recent spectra, which in fact did not show relevant changes in the line and continuum flux level (Costa \& de Freitas Pacheco 1994, Pereira 1995, Pereira et al. 1998).
\section{The theoretical framework}
\subsection{The colliding-wind scenario}
In the past years, theoretical models (Girard \& Willson 1987, Kenny \& Taylor 2005)
as well as observations (Nussbaumer et al. 1995) have categorically shown that in SS both the hot and cool stars lose mass through strong stellar winds which collide within and outside the system, hence creating a complex network of wakes and shock fronts which result in a complicated structure of gas and dust nebul\ae{} (Nussbaumer 2000).\\
In this paper, as previously done for other SS (see Sect. 3.2), we refer to two main shocks: the shock between the stars facing the WD, which is a head-on shock (hereafter the \textit{reverse} shock), and the head-on-back shock, which propagates outwards the system (hereafter the \textit{expanding} shock). Both the nebulae downstream of the shock fronts are ionized and heated by the radiation flux from the hot star and by shocks. The photoionizing radiation flux reaches the very shock front of the reverse shock, while it reaches the edge opposite to the shock front downstream of the expanding shock.
\subsection{The calculation code}
The results presented in this work are performed by SUMA (Viegas \& Contini 1994; Contini 1997), a numerical code that simulates the physical conditions of an emitting gaseous cloud under the coupled effect of ionization from an external radiation source and shocks, and in which both line and continuum emission from gas are calculated consistently with dust reprocessed radiation (grain heating and sputtering processes are also included). The derived models have been successfully applied to several SS, e.g. AG Peg (Contini 1997, 2003), HM Sge (Formiggini, Contini \& Leibowitz 1995), RR Tel (Contini \& Formiggini 1999), He2-104 (Contini \& Formiggini 2001), R Aqr (Contini \& Formiggini 2003), HD330036 (Angeloni et al. 2007b), as well as to nova stars (V1974 Cyg, Contini et al. 1997 - T Pyx, Contini \& Prialnik 1997) and supernova remnants (Kepler's SNR, Contini 2004).\\

The calculations start with gas and dust entering the shock front in a steady state regime: the gas is adiabatically compressed and thermalized throughout the shock front. In the downstream region the compression is derived by solving the Rankine-Hugoniot equations (Cox 1972): the downstream region is automatically divided in plane parallel slabs in order to calculate as smoothly as possible the physical conditions throughout the nebula. Radiation transfer and optical depths of both continuum and lines are calculated for a steady state: in particular, radiation transfer of the diffused radiation is taken into account following Williams (1967). The fractional abundance of the ions in different ionization stages is calculated in each slab by solving the ionization equilibrium equations for the elements H, He, C, N, O, Ne, Mg, Si, S, Cl, Ar, and Fe.
The electron temperature in each slab is obtained from the energy equation when collisional processes prevail and by thermal balancing when radiation processes dominate.\\
Compression downstream strongly affects the gas cooling rate by free-free, free-bound, and line emission: consequently, the emitting gas will have different physical conditions depending on the shock velocity and on the pre-shock density. For example, self-absorption of free-free radiation, consistently calculated in every slab, affects the emission spectral index at low radio frequencies (see Sect. 4.2.1)

Dust is included in the calculations, too. Dust and gas are coupled throughout the shock-front and downstream by the magnetic field. In each slab the sputtering of the grains is calculated, leading to grain sizes which depend on the shock velocity and on the gas density. The temperature of the grains, which depends on the grain radius, is calculated by radiation heating from the external (primary) source and by diffuse (secondary) radiation, as well as by gas collisional heating. The dust reprocessed radiation flux is calculated by the Plank-averaged absorption coefficient of dust in each slab, and integrated throughout the nebula downstream. 

The input parameters which characterise the shock are the shock velocity, \textit{\Vs}, the
preshock density of the gas, \textit{\n0}, and the preshock magnetic field, \textit{\B0}. The radiation
flux is determined by the temperature of the star, interpreted as a colour temperature, \textit{\Ts},
and by the ionization parameter, \textit{U}. The dust-to-gas ratio, \textit{$d/g$} is also accounted for, as well as the relative abundances of the elements to H. 

A full detailed description of the code is to be presented in Contini \& Viegas (2007 in preparation).
\begin{figure*}[!ht]
\begin{center}
\includegraphics[width=0.46\textwidth]{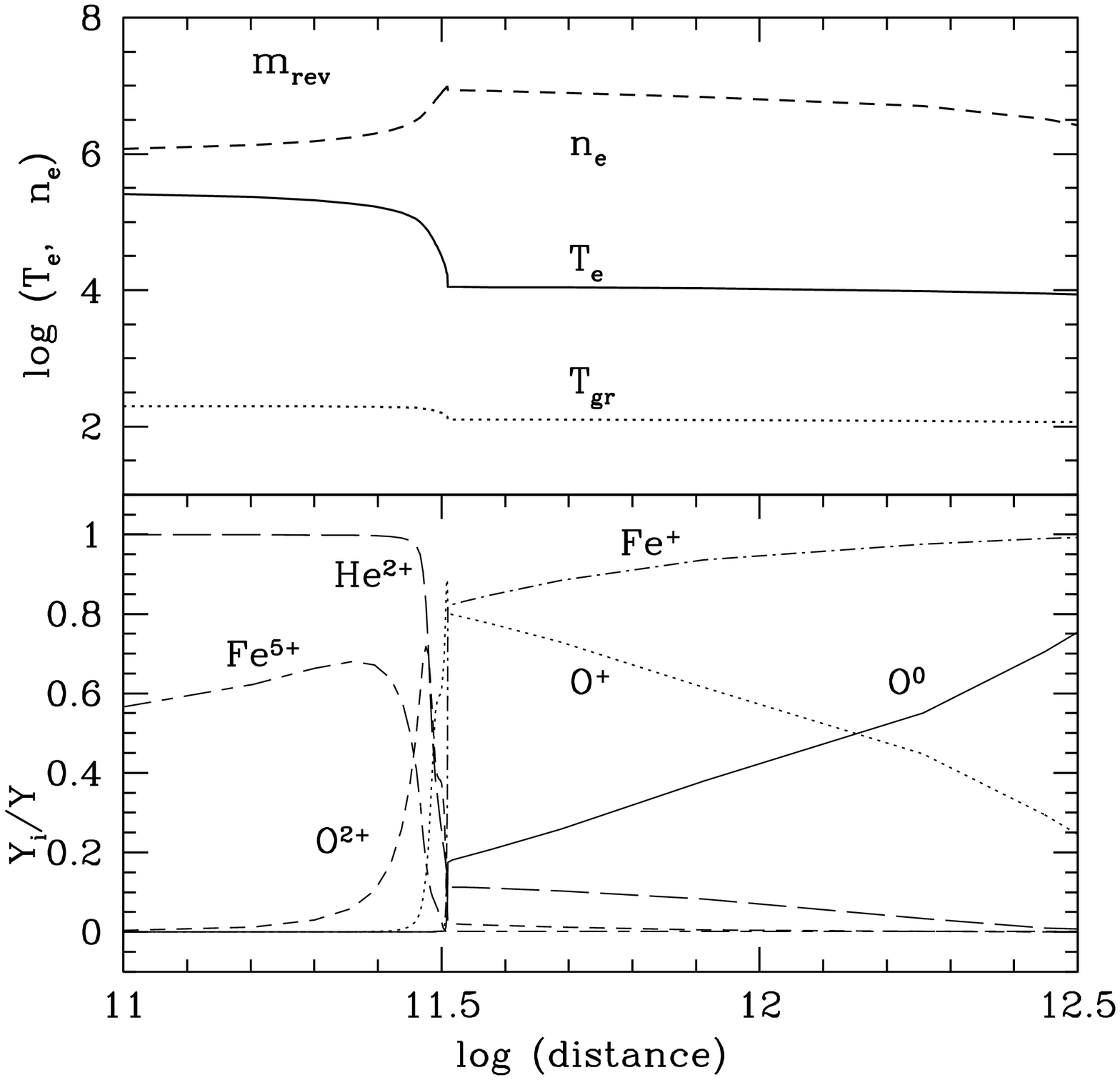}
\includegraphics[width=0.46\textwidth]{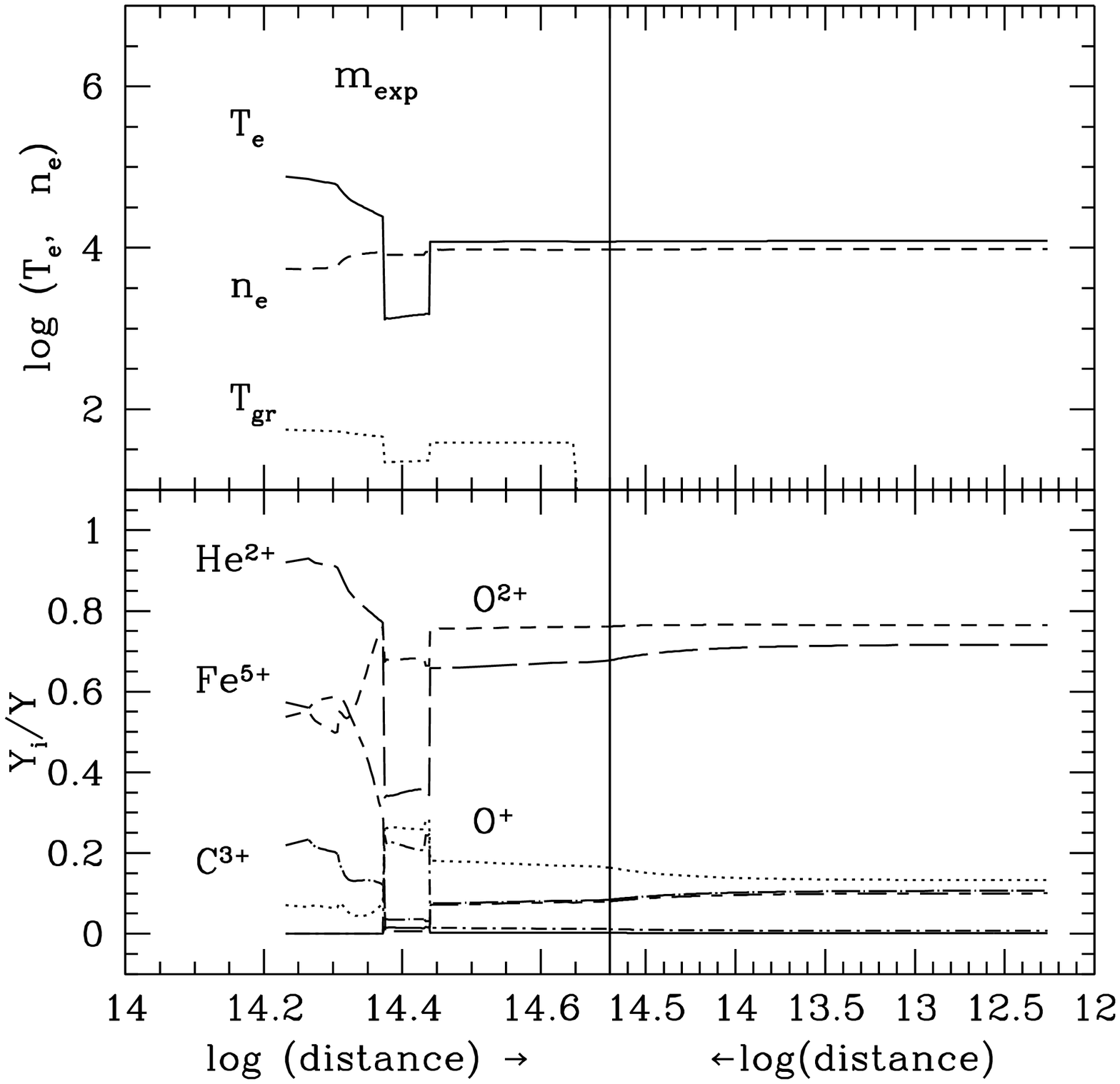}
\caption{The profile of the electron temperature, of the electron density and of the temperature of the grains (\textit{\agr}=0.2 \mum) downstream of the reverse (left diagram) and
of the expanding shock front (right diagram). The stratification of the most significant ions is shown in the bottom panels. For both diagrams, shock front is on the left side.\label{fig:strat}}
\end{center}
\end{figure*}
\begin{table}[!ht]
\centering \caption{The radio-mm data references. \label{tab:radioref}}
\begin{tabular}{ccc}\\ 
\hline  \hline
$\nu$ [GHz] & Flux [mJy] & Ref.\\
\hline
0.843 & 10.6$\pm$1.2 & Jones (1985)\\
1.4	& 13.5 & Siodmiak \& Tylenda (2001)\\
1.6	& 15 & Taylor \& Seaquist (1984)\\
2.7	& 33$\pm$4 & Purton et al. (1982)\\
4.885 & 50 & Aaquist \& Kwok (1990)\\
5.0	& 46$\pm$10 & Purton et al. (1982)\\
5.0	& 50 & Siodmiak \& Tylenda (2001)\\
6.2	& 65$\pm$10 & Purton et al. (1982)\\
8.1	& 82$\pm$7 & Purton et al. (1982)\\
8.31 & 73.5 & Bastian (1992)\\
8.44	& 65.3$\pm$3.3 & Seaquist et al. (1993)\\
8.9	& 90$\pm$15 & Purton et al. (1982)\\
14.5	& 90$\pm$10 & Purton et al. (1982)\\
14.965  & 72 & Aaquist \& Kwok (1991)\\
90.0	& 127$\pm$53 & Purton et al. (1982)\\
231 & 55$\pm$16 & Ivison et al. (1995)\\
\hline
\end{tabular}
\end{table}
\begin{table*}[!ht]
\centering \caption{The infrared data references.\label{tab:irref}}
\begin{tabular}{cccccccccccccccccc}\\ 
\hline  \hline
$\lambda$ [\mum] (phot. band) & Flux [Jy]$^a$ & Ref.\\
\hline
1.20 (J), 1.64 (H), 2.22 (K), 3.8 (L') & 1.55 - 2.1 - 4.8 - 7.1 & Allen (1983)\\ 
8.28, 12.13, 14.65 & 7.454 - 11.93 - 14.67 & MSX6C PSC, Egan et al.$^b$ (2003)\\
1.235 (J*), 1.662 (H*), 2.159 (K*) & 0.0312 - 0.115 - 0.594 & Ramos-Larios \& Phillips$^c$ (2005)\\
0.9 (I$_c$), 1.25 (J), 2.22 (K) & 0.0167 - 0.034 - 0.741 & DENISE$^d$\\
12, 25, 60 & 18.13 - 28.16 - 5.75 & IRAS PSC$^e$\\
34.6 & 16.6 & He et al. (2005)$^f$\\
2.5-45.5 SWS &- - -&ISO IDA$^g$\\
\hline
\end{tabular}
\flushleft
\begin{tiny}
* 2MASS Photometric System;
a: out of Allen, expressed in 10$^{14}$ erg s$^{-1}$ cm$^{-2}$ A$^{-1}$;
b: The Midcourse Space Experiment Point Source Catalog Version 2.3 (October 2003);
c: 2MASS NIR measurements of galactic planetary nebulae;
d: The DENIS consortium (third release - Sept.2005);
e: IRAS Catalogue of Point Sources, Version 2.0 (IPAC 1986);
f: The 35 \mum\ absorption line towards 1612 MHz masers;
g: ISO Data Archive (@ http://www.iso.vilspa.esa.es).
\end{tiny}
\end{table*}
\section{Modelling the spectra}
In this section we present the models which best reproduce the whole properties of H1-36. The models dealing with the observed line spectra appear in Sect. 4.1. They are cross-checked by the continuum SED (Sect. 4.2) until a fine tuning of line and continuum spectra is found: we refer in particular to the SED in the radio range (Sect. 4.2.1) where the turnover frequency and the steeper spectral index of the bremsstrahlung at lower frequencies are highly sensitive to the physical conditions of the gas.
\subsection{The line ratios}
The spectra derived from Allen's observations are shown in Table \ref{tab:mod}, col. 2; the models calculated for the reverse ($m_{rev}$) and for the expanding shock ($m_{exp}$) are presented in cols. 3 and 4, respectively; their weighted sum appears in col. 5. 

The input parameters of the models appear in Table \ref{tab:modpar}. They are constrained by the observations, namely, the shock velocity \textit{\Vs}\ is roughly determined by the FWHM of the line profiles (in agreement with Allen 1983); the preshock density, \textit{\n0}, is constrained by the slope of the radio continuum and by the ratios of characteristic lines (e.g. [SII] 6717/6731); the preshock magnetic field, \textit{\B0}, is characteristic of symbiotic systems (Crocker et al. 2001); the stellar temperature, \Ts, is taken from the analysis of Allen (1983); the ionization parameter, \textit{$U$}, is determined phenomenologically; the geometrical thickness, D, is constrained by the radio continuum slope and by the ratios of the lines to \Hb; eventually, the relative abundances of the elements are determined by the line ratios. Unfortunately, in the present case, the dust-to-gas ratio \textit{(d/g)} cannot be directly determined because the bump of the dust reprocessed radiation is covered by the emission from the summed dust shell and Mira black body fluxes.

The line ratios calculated by $m_{rev}$ and $m_{exp}$ are summed up adopting the weights \textit{w} shown in Table \ref{tab:modpar}, last row. They are the same as those adopted in Fig. \ref{fig:sed} to suite the continuum SED. 

The observed line ratios are reproduced by the summed model within a factor of 2, except a few lines which are badly fitted, e.g. [NI] 5200, since they have a particularly low critical density for collisional deexcitation and are very faint at the relatively high density of $m_{rev}$ and $m_{exp}$. We suggest that the observed  HeII 1640/ \Hb\ may be blended with the rather high OIII 1640/\Hb\ ratio, whose calculated value is $>$ 3. An improved fit of the [OIII] 5007/[OIII] 4363 ratio could result from a reverse shock with lower densities, which are however incompatible with the radio SED: as a matter of fact the [OII] 3727 and [OIII] 5007 lines are mostly emitted from the expanding shock.

Model $m_{rev}$ shows N/H and O/H slightly higher than solar by factors of 1.7 and 1.3 respectively, while Fe is strongly depleted both in $m_{rev}$ by a factor of 10 and in $m_{exp}$ by a factor of 5: this is not surprising as iron, a refractory element, is easily trapped into dust grains.

In Fig. \ref{fig:strat} we present the profile of the electron density, electron temperature and the fractional abundances of the most significant ions throughout the nebula downstream of the reverse (left diagram) and expanding shock  (right diagram). It is worth noticing that model $m_{rev}$ is matter-bound: this is characteristic of the interbinary symbiotic nebulae, where the geometrical thickness is constricted by the colliding wind region.
\begin{table}[!ht]
\centering \caption{The fluxes of the UV and optical spectral lines, normalized to \Hb.  \label{tab:mod}}
\begin{scriptsize}
\begin{tabular}{ccccccccc}\\ \hline  \hline
line & obs$^a$&  $m_{rev}$ & $m_{exp}$ & av$_{m_{rev}+m_{exp}}$\\
\hline
\ CIV 1549 &18.78& 19.7   & 27.3& 22        \\
\ HeII 1640& 7.88$^b$  &1.1  & 5.2 &2.6      \\
\ NIII] 1749 &5.15& 2.24 & 2.  & 2.2       \\
\ CIII] 1909 &27.2& 9.3 & 18. & 13.0    \\
\ [NeV]+OIII 3425 & 4.24 &0.13$^c$ & 0.13$^c$ &0.22$^c$ \\
\ [FeVI] 3662&0.006& 0.0043&0.0047 &0.0044\\
\ [OII]+ 3727&0.576& 0.04 & 2.6 & 0.99\\
\ [NeIII] 3869+ &2.12& 1.5 & 2.2  & 1.78\\
\ [SII]+ 4070&0.18& 0.38 & 0.08 & 0.27\\
\ H$\gamma$ & 0.42& 0.44 & 0.45 &0.44 \\
\ [OIII] 4363 & 0.61& 1.29 & 0.80 & 1.1 \\
\ HeI 4471     &0.056& 0.036 & 0.023 & 0.044\\
\ [FeIII] 4658 & 0.015& 0.01  & 0.015 & 0.012\\
\ HeII 4686  & 0.55& 0.16 & 0.76 & 0.4 \\
\ \Hb\  4861  & 1  & 1   &  1 &  1\\
\ [ArIV] 4711  &0.02&0.0013 & 0.13 & 0.049\\
\ [NeIV]+ 4714 & 0.11& 0.14 & 0.08 & 0.12\\
\ [ArIV] 4740  & 0.097&0.011 & 0.15 & 0.063\\
\ [OIII[ 5007+ & 19.1&  2. & 35. & 14.2\\
\ [FeVI] 5146  &0.017&0.007 & 0.03 & 0.016\\
\ [FeII] 5159 &0.03& 0.039 & 0.0017 & 0.025 \\
\ [FeVI] 5176  & 0.03& 0.018 & 0.033 & 0.023\\
\ [ArIII]5191 & 0.005& 0.017 & 0.012 & 0.015\\
\ [NI] 5199   & 0.006& 0.0002  & 0.0005 & 0.0003\\
\ [FeVI] 5335 &0.011& 0.0018 & 0.017 & 0.008 \\
\ [NII] 5755 &0.15&0.43& 0.04 & 0.28\\
\ HeI 5876   &0.11& 0.18 & 0.07 & 0.14 \\
\ [FeVII] 6087& 0.12&0.002  & 0.005 & 0.003\\
\ [OI] 6300+  & 0.4&0.47&0.013 & 0.3 \\
\ [SIII] 6312 & 0.08&0.14 &0.13 & 0.136\\
\ [NII] 6548+ & 1.0& 0.2 & 1.4 & 0.63\\
\ H$\alpha$  & 4.5& 3.5 & 3.1 & 3.4 \\
\ [SII] 6717 & 0.02&0.002 & 0.057 & 0.022\\
\ [SII] 6731 & 0.039&0.0035& 0.10 & 0.039\\
\ [ArIII] 7136 & 0.24&0.14 & 0.24 &0.14\\
\ [ArIV] 7170  & 0.007&0.023& 0.009 & 0.018\\
\ [OII] 7319+ &0.55& 2.6  & 0.48 & 1.8\\
\ [FeII] 8617 & 0.0027&0.0027 & 0.0011 & 0.002 \\
\ [SIII] 9069 & 0.33&0.06& 1.3 & 0.5\\
&&&\\
\Hb\ abs.$^d$ & - &1.13 &0.0075&-\\
\hline
\end{tabular}
\begin{flushleft}
a: observed values from Allen (1983);\\
b: blended with OIII 1640;\\
c: the calculated flux refers only to [NeV];\\
d: \Hb\ absolute flux calculated at the nebul\ae{}. [\erg].
\end{flushleft}
\end{scriptsize}
\end{table}
\begin{table}[!hb]
\centering \caption{The model input parameters. \label{tab:modpar}}
\begin{tabular}{cccccccccccccc}\\ \hline  \hline
parameter & $m_{rev}$ & $m_{exp}$ \\
\hline
\  \Vs\ (\kms) & 140 & 70  \\
\  \n0\ (\cm3) &2.5e5 & 3.5e3 \\\  
\  \B0\ (gauss) & 1e-3& 1e-3 \\
\  \Ts\ (K)     & 1.5e5 & 1.5e5  \\
\  U           & 2e-3  & 2.5e-3 \\
\  D (cm)      & 2.8e12 & 1e15 \\
\  $d/g$      &  1e-14 & 1e-14 \\
\  He/H       &  0.1   & 0.1 \\
\  C/H        & 3.3e-4 & 3.3e-4\\
\  N/H        &1.5e-4  &9.1e-5\\   
\  O/H        & 8.6e-4  &6.6e-4\\
\  Ne/H       & 8.3e-5 &8.3e-5\\
\  Mg/H       &2.6e-5  &2.6e-5\\
\  Si/H       &3.3e-5 &3.3e-5\\
\  S/H        & 1.6e-5 & 1.6e-5\\
\  Ar/H       &6.3e-6 & 6.3e-6\\
\  Fe/H       & 3.2e-6 &6.2e-6\\
\  log $w$ & -11.05 & -9.1 \\
\hline
\end{tabular}
\end{table}

\subsection{The continuum SED}
\begin{figure}[!hb]
\begin{center}
\includegraphics[width=0.48\textwidth]{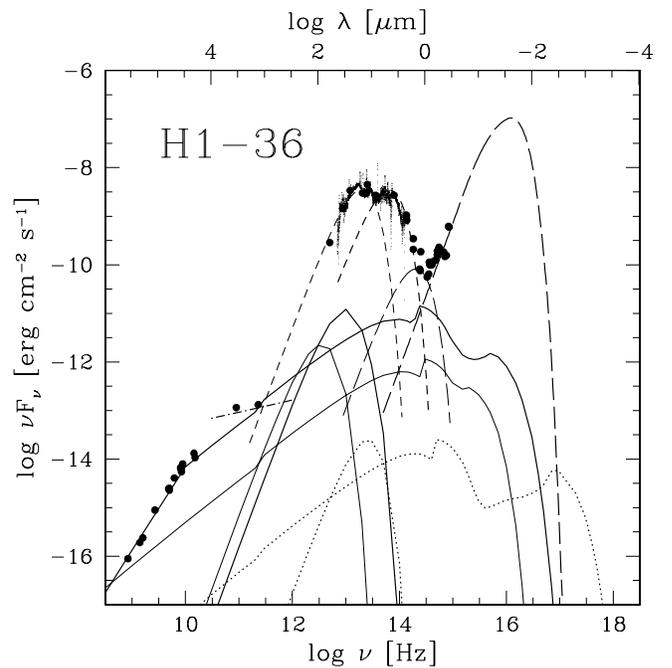}
\caption{The H1-36 continuum SED, from radio to UV. Thin short-dashed line: 250 K dust shell; thick short-dashed line: 800 K dust shell; thin long-dashed line: Mira (2500 K) stellar component; thick long-dashed line: hot stellar component; dot-dashed line: synchrotron; thin solid line: $m_{exp}$; thick solid line: $m_{rev}$ (see Table \ref{tab:modpar} for details about the model input parameters); dotted line: the \textit{jet} (see Sect. 4.3). Notice that for $m_{rev}$, $m_{exp}$ and the \textit{jet} models there are two curves: one is the bremsstrahlung emission from radio to X-rays, the other represents the consistently calculated reprocessed radiation by dust in the IR.
\label{fig:sed}}
\end{center}
\end{figure}

In Fig. \ref{fig:sed} we present the modelling of the continuum. The SED is the result of the emitting contributions from the cool and hot stars, as well as of the fluxes from the dust shells and of the bremsstrahlung from the shocked nebulae downstream of the shock fronts, which emit the UV and optical line spectra. \\
The bremsstrahlung from the nebul\ae{} shows two main peaks. At higher frequencies the continuum is emitted from gas collisionally heated by the shock at relatively high temperatures: the peak frequency depends therefore on the shock velocity. The peak at $\sim$ 10$^{14}$Hz, on the other hand, depends on the volume of gas at temperatures of $\sim$ 1-3 10$^4$ K, which is heated and ionized mainly by the photoionizing flux: the peak frequency is therefore more sensitive to the radiation parameters, \Ts\ and \textit{$U$}. Also a contribution to the SED by synchrotron radiation, not so exotic in objects where shocks are at work, can not be ruled out (see Sect. 4.2.1 and Fig \ref{fig:sed}). \\
As a matter of fact, without more constraining data it remains puzzling to determine whether the measured flux densities at $\sim$ 10$^{11}$Hz are dominated by synchrotron, by bremsstrahlung or by thermal emission from cold dust, also taking into account that emission from either mechanism could fluctuate in response to variable mass-loss episodes.
\subsubsection{The radio continuum}
In an important paper on radio emission of mass-losing stars, Wright \& Barlow's (1975) presented an analytic study of the spectral flux distribution produced by completely ionized, uniform, spherically symmetric mass loss flow in early type star. Though under idealised hypothesis, they tried to interpret the radio data for some recently observed objects in the light of thermal free-free emission which would result in a power-law spectrum $S_{\nu} \propto \nu^{0.6}$. They also discussed deviations from the ideal treatment caused by e.g., non-uniform mass-loss rates or by the actual ionization structure of the emitting circumstellar envelope.

In effect, the Purton's radio survey (Purton 1982) pointed out that there were some objects (with a high incidence of SS) whose radio spectral index was systematically steeper than $\alpha$=0.6 (in the range 0.8-1.5), therefore implying an envelope density profile quite far from the inverse-square one assumed by Wright \& Barlow (1975). In any event, no attempt was made by Purton to fit a detailed physical model to the data.

Few years later, Seaquist et al. (1984) and Taylor \& Seaquist (1984) presented a model (hereafter STB model) which explained the H1-36 radio properties in light of the mounting evidence for the binary nature of symbiotic stars: by assuming that the emitting region arises from the portion of the cool component wind photoionized by the WD, they performed a fit to the Purton's data with a least-square criterion as a function of the so called X parameter (related to the shape of the ionized nebula), of the turnover frequency $\nu_t$, and of the flux density at $\nu_t$. The model was successful in reproducing many of the observed properties of radio symbiotics, such as the spectral index, the spectral turnover, the correlation between the radio emission and the spectral type of the cool component. Nonetheless, it was soon recognised that the actual radio sources were more complex than those assumed in the STB model, and that any derived physical property had to be treated with particular caution. Furthermore, more recent observations by Seaquist \& Taylor (1993) at millimeter and submillimeter wavelengths showed that several objects might differ considerably from the STB model.\\ As with regards to the specific case of H1-36, the STB model failed e.g. to reproduce the 843MHz flux presented by Jones (1985). Moreover, the assumed density profile appeared too simplistic: even the two-layer model discussed by Costa \& de Freitas Pacheco (1994) referring to the inner structure of the emitting nebulae cannot be considered adequate since SS have such important density gradients that do not even allow to adopt unique values for electron temperature and densities.\\
Unfortunately, so far there are no papers which try to interpret the radio spectra taking into account the possibility of any outflow from the compact star.

In the following we fit the existing radio-mm data we have presented in Table \ref{tab:irref} and Fig. \ref{fig:var} - bottom panel, within the colliding-wind scenario depicted in Sect. 3.1. Our approach is shown to be self-consistent with the continuum profile of the system along the whole electromagnetic range, as well as cross-checked by the spectral line emission.

Three main slopes are seen between 10$^8$ and 10$^{11}$ Hz. \\
The flattest one, with a spectral index of 0.75, can be explained by synchrotron radiation created by the Fermi mechanism at the shock front (Bell 1978a,b): a cut-off at the lower limit is given by Ginzburg \& Syrovatskii (1965) ($\nu$ $\sim$ 20 N$_e$/B, with N$_e\sim$10$^6$ \cm3\ and B=10$^{-3}$ gauss at the shock front), explaining the SED profile at $\sim$ 10$^{11}$Hz. The unabsorbed bremsstrahlung explains the slope of the SED between about 10$^{10}$ and 10$^{11}$ Hz. At $\nu<10^{10}$ Hz the bremsstrahlung shows the high slopes which are produced by free-free self-absorption. The optical depth $\tau=8.24 \, 10^{-2}T^{-1.35}\nu^{-2.1}E$ (where T is to be measured in K, $\nu$ in GHz and E, the so called \textit{emission measure} $E=\int n_e n^+ ds$, in cm$^{-6} pc$ - Osterbrock 1988) is low for gas at relatively high temperatures: however, in the downstream nebulae of H1-36 we found large regions (i.e. large $ds$) of gas at T $\sim$ 10$^4$ K (Fig. \ref{fig:strat}) which make $\tau > 1$. The other parameters which vary from slab to slab are the density, which is calculated through the compression equation, and the geometric thickness of the slab. At this point we would like to emphasise that the calculation of the density profile is not assumed \textit{a priori} in our models accounting for shocks, but consistently calculated across the whole emitting nebula (Sect. 3.2). In this case, the thinner the slab the better is the approximation of the calculated spectra. Therefore the densities must be sufficiently high to give a sensible absorption. The turnover corresponds to different frequencies for different slabs and the final bremsstrahlung results from the integrated flux throughout the whole length of the nebula. The actual turnover frequency $\nu_t$ indicates then the transition from the optically thin to the optically thick case.

Eventually, it is worth highlighting that higher densities generally refer to the nebula downstream of the shock front facing the WD between the stars (model m$_{rev}$), while the densities downstream of the expanding shock (model m$_{exp}$) correspond to an optically thin nebula. Therefore, the radio spectrum observed in SS may be emitted by the gas between the stars, with an important, more increasing contribution from the expanding nebula at lower frequencies. Interestingly, this agrees with the variable size of the radio nebula, which becomes gradually larger at lower frequencies (i.e. 0.6 arcsec at $\lambda$=2 cm and 5 arcsec at $\lambda$=6 cm - Taylor 1988).
\subsubsection{The IR continuum}
One of the most intriguing aspect of the symbiotic phenomenon in D types pertains to the dusty environment. As a matter of fact, these systems show a broad IR excess which has been attributed to emission from circumstellar dust since the first IR surveys. Early observations already showed that in D type objects the dust excesses have colour temperatures near 1000 K (Feast et al. 1983); nevertheless today, thanks to the IR astronomy developments, it has been realized that it is difficult to explain the observed IR spectrum by a single temperature. Theoretical models too, confirmed that different ''dust'' temperatures should be combined in order to reproduce the NIR-MIR data (e.g. Anandarao et al. 1988, Schild et al. 2001, Angeloni et al. 2007c, in preparation).

In H1-36, the temperature of the dust shells is derived by modelling the IR data we collected which, quite surprisingly, agree in the overlapping frequency range: this proves that variability, even if present as in all symbiotic IR spectra, has not been as large as to substantially modify the dust continuum emission and then invalidate our results (see Sect. 2, Fig. \ref{fig:var} and Table \ref{tab:irref}). \\
It is worth noticing that the cooling rate is very strong at the densities typical of the region closer to the shock front downstream, which corresponds to the maximum temperature of dust: this implies that the flux from the shell corresponds mainly to the maximum temperature of the grains and can be modelled by a black body (bb). As a matter of fact, Fig. \ref{fig:sed} shows that the whole IR continuum can be well explained by the combination of two bb curves, corresponding to temperatures of 800 K and 250 K.
By comparing the models with the data we obtain the $\eta$ factors: they depend on the radius of the dust shell, $r$, and on the distance of the system to Earth $d$ ~ ($\eta=r^2/d^2$), being the fluxes calculated at the nebula and the data, obviously, measured at Earth. Adopting \textit{d}=4.5 kpc (Allen 1983), we find \textit{r}= 4.4 10$^{14}$ cm  and 4.7 10$^{15}$ cm for the shells at 800 K and 250 K, respectively. Both the derived inner shell temperature and radius are in good agreement with what found by Allen (1983), who suggested that the cool component is heavily embedded in its own dust shell, specifically the hot shell is circumstellar.\\
An interesting issue is related to the outer shells: if we relied on the Allen's proposed binary separation (3 10$^{16}$ cm), then even this shell would be circumstellar, namely, well comprised within the system. Allen himself admitted such a large binary separation was striking, though; moreover, he stated that, translated in an angular separation of $\sim$0.5 arcsec according to the suggested distance of 4.5 kpc, this binary separation would have been large enough to be resolved by careful infrared astrometry, as well as by high-resolution optical and radio observations. As a matter of fact, few year later Taylor (1988) succeed in resolving the radio nebula. Conversely, Bhatt \& Sagar (1991) failed in detecting any nebulosity in the optical emission lines, arguing that if this was actually present, it should be smaller than $\sim$2 arcsec. Corradi et al. (1999) finally resolved the optical nebula and found out that it appeared also in a continuum image, therefore suggesting that it might be a reflection nebula.

At this point, we would like to stress that the unusual large size of the system has been constrained so far by observations of gaseous emitting nebul\ae{}, and that the binary stellar components have never been resolved: this might be a clue that the actual binary separation is overestimated, therefore implying that the outer shell we found directly by modelling the IR data may even be circumbinary, surrounding both the stars. This would not be so out of ordinary for D-type SS, since this dichotomy in the dust shell distribution (with a hot shell being circumstellar, and a cool one circumbinary) has been confirmed by recent studies (Angeloni et al. 2007c, in preparation).

H1-36 appears as an extreme SS: only further observations at higher spatial resolution and in different spectral ranges will be able to investigate the actual size and physical properties of the dust and gas nebul\ae{}.
\subsubsection{The UV continuum}
For sake of completeness we would like to note that longward of $log\ \nu$=14.5 the UV data are well fitted by a \Ts=150.000 K bb curve (Fig. \ref{fig:sed}), in agreement with the Allen's (1983) derived hot star temperature: the WD flux dominates in that range over the bremsstrahlung emission. However, it is worth reminding that between $log\ \nu \sim$ 16.4 and $log\ \nu \sim$ 17.7 absorption of X-rays is quite strong, therefore implying that X-rays, if present, would be soft and mainly due to bremsstrahlung emission arising from the high-velocity component we deduce from the IR line features (see Sect. 4.3).
\subsection{A jet in H1-36?}
In a previous analysis of ISO SS infrared spectra (Angeloni et al. 2007a - hereafter Paper I), we found broad emission lines, indicating high-velocity components in many objects of the sample (Paper I - Fig. 1, Table 4). In H1-36, the FWHM of these line profiles correspond to velocities of 500 \kms; we remind that high-velocity components were not observed in the optical-UV spectra presented by Allen (1983) which showed FWHM $\leq$ 200 \kms. Consequently, since the ISO-SWS spectrum shows several highly-ionized and broad Ne and O lines, not compatible with the physical models derived by the optical-UV spectra, we decide to include these new results in order to draw out an as most large and consistent as possible interpretation. We then calculated a further model, characterised by higher shock velocities and lower densities, able to reproduce the observed IR lines without making the previous fit worse.

In Paper I we found that, in H1-36, the [NeVI] 7.65 \mum\ line dominates, followed by the [OIV] 25.89 \mum, [NeIII] 15.55 \mum, [NeV] 24.31 \mum, and [NeV] 14.32 \mum\ lines. \\
The model calculated with \textit{\Vs}=500 \kms, \textit{\n0}=10$^4$ \cm3, \textit{$U$}=8, and D=10$^{16}$ cm well reproduces the observed intensity ratio: in fact we obtain [NeII]12.8/[NeVI]7.65=0, [NeV]14.32/[NeVI]7.65=0.47, [NeIII]15.55/[NeVI]7.65=1.3e-4, and [NeV]24.31/[NeVI]7.65=0.05, in agreement with the observed upper limits, respectively [NeV]14.32/[NeVI]7.65$<$0.5, [NeIII]15.55/[NeVI]$<$0.2 and [NeV]24.31/[NeVI]$<$0.15. 
Moreover, the same model represents a non negligible contribution only on the high ionization level lines, particularly to CIV 1549, [NeV] 3425, [FeII]6087, and a relatively strong contribution to HeII 1640.

We can now calculate the distance of this high-velocity structure from the WD. A WD \Ts=150.000 K corresponds to a ionizing photon flux of 4 10$^{26}$ photons cm$^{-2}$ s$^{-1}$: this flux is related to the ionization parameter and to the gas number density, in the radiation dominated zone, by $F_{\nu}(R_{WD}/r_j)^2 = U n c $. Adopting an average density of n=10$^5$ \cm3 after compression, and a WD radius of 10$^9$ cm, we obtain log $\eta$ = -16.2. This value allows to place the bremsstrahlung emission from this component in the continuum SED diagram of Fig. \ref{fig:sed}, and to verify that its contribution to the radio spectrum is negligible when compared to the bremsstrahlung from the nebul\ae{}. Conversely, on the high-energy side of the spectrum, it is interesting to notice that \textit{\Vs}=500 \kms\ would correspond to a temperature in the immediate postshock region of $\sim$3.6 10$^6$ K, hence indicating that a soft X-ray emission may be  actually present in H1-36.

Eventually, after this consistent analysis, we would like to stress the possibility that this high-velocity component in H1-36 is a jet-like feature. This suggestion, if confirmed, would be fascinating, as only recently it has been recognising that SS are a class of jet-producing objects. Furthermore, the two SS X-ray jets that have been discovered to date (R Aqr, Nichols et al. 2007 and CH Cyg, Karovska et al. 2007) both show a X-ray structure more extended than their radio ones (Sokoloski et al. 2006). This would explain why this 10$^{16}$ cm jet has not been seen in the radio image (with comparable angular resolution) and drive us to indicate SS as promising X-ray targets.
\section{Concluding remarks}
In this paper we have analysed the spectra of D-type SS H1-36 within a colliding-wind theoretical framework. After having constrained the relative physical conditions by modelling more than 40 emission lines from radio to UV, we have been able to confidently explain the continuum SED by taking into account all the emitting contributions arising from both the stars, the dust shells and the gaseous nebul\ae{}. A comprehensive model of the radio spectra allowed to reproduce the different slopes of the radio profile and the turnover frequency, as well as the size of the observed nebul\ae{} at several frequencies in light of different contribution from the expanding and reverse shocks. The IR continuum unveiled the presence of two dust shells with characteristic radii and temperatures: the inner shell is confirmed to embed the Mira star, while the outer one may be \textit{circumbinary}, i.e. surrounds the whole binary system. We believe that the presence of multiple dust shells is not a unique characteristic of H1-36, but it may somehow represent a distinctive feature of D-type SS. Furthermore, the broad profiles of the IR lines directed us to investigate whether a high-velocity component (perhaps an X-ray jet) may be present. This insight led us to indicate H1-36 as a promising X-ray target and represents a further support to the emerging interpretation of SS as a class of jet-producing objects.
We then encourage new observations and studies which consistently take into account the complex nature of SS throughout the whole electromagnetic spectrum.
\begin{acknowledgements}
The authors are very grateful to the anonymous referee for many helpful comments that improved the readability of the paper.
RA acknowledges the kind hospitality of the School of Physics \& Astronomy of the Tel Aviv University.
\end{acknowledgements}

\end{document}